\documentclass[times,twocolumn,final]{elsarticle}

\usepackage{cag}
\usepackage{framed,multirow}
\usepackage{amsmath}

\usepackage{amssymb}
\usepackage{latexsym}

\usepackage{url}
\usepackage{xcolor}
\definecolor{newcolor}{rgb}{.8,.349,.1}

\usepackage{hyperref}

\usepackage{bbm}
\usepackage{bm}
\usepackage{xspace}
\usepackage{algorithm2e}
\usepackage{algorithmicx}
\usepackage[hang,nooneline]{subfigure}
\usepackage{cases}

\newcommand{\Bezier}{B\'ezier\xspace}
\newcommand{\Bezierblock}{B\'ezier block\xspace}
\newcommand{\block}{block\xspace}

\usepackage{amssymb}
\newcommand{\R}{\ensuremath{\mathbb{R}}}
\newcommand{\N}{\ensuremath{\mathbb{N}}}
\newcommand{\glow}{\sigma}
\newcommand{\reffig}[1]{\figurename~\ref{fig:#1}}
\newcommand{\refequ}[1]{Equation~\eqref{eq:#1}}
\newcommand{\refsec}[1]{Section~\ref{sec:#1}}

\newcommand{\refalg}[1]{Algorithm~\ref{alg:#1}}

\newcommand{\locRefSpl}{LR-spline\xspace}

\newcommand{\EC}[1]{}

\usepackage{tikz}
\usetikzlibrary[graphs, positioning, arrows]
\newcommand{\drawLRelements}[1]{
        \draw[gray,thick] (0+#1,0) rectangle (4+#1,4);
        \draw[gray,thick,-] (1+#1,0) -- (1+#1,4);
        \draw[gray,thick,-] (1+#1,3) -- (4+#1,3);
        \draw[gray,thick,-] (2+#1,3) -- (2+#1,4);
        \draw[gray,thick,-] (2+#1,3.5) -- (4+#1,3.5);
        \draw[gray,thick,-] (3+#1,3.5) -- (3+#1,4);

        \def\dep{2}
        \draw[gray,thick,-] (0+#1,4) -- (#1+\dep,4+\dep);
        \draw[gray,thick,-] (1+#1,4) -- (1+#1+0.25*\dep,4+0.25*\dep);
        \draw[gray,thick,-] (2+#1,4) -- (2+#1+\dep,4+\dep);
        \draw[gray,thick,-] (3+#1,4) -- (3+#1+\dep,4+\dep);

        \draw[gray,thick,-] (4+#1,4) -- (4+#1+\dep,4+\dep);
        \draw[gray,thick,-] (4+#1,3.5) -- (4+#1+0.25*\dep,3.5+0.25*\dep);
        \draw[gray,thick,-] (4+#1,3) -- (4+#1+\dep,3+\dep);
        \draw[gray,thick,-] (4+#1,0) -- (4+#1+\dep,0+\dep);

        \draw[gray,thick,-] (#1+\dep,4+\dep) -- (4+#1+\dep,4+\dep);
        \draw[gray,thick,-] (#1+0.25*\dep,4+0.25*\dep) -- (4+#1+0.25*\dep,4+0.25*\dep);
        \draw[gray,thick,-] (3+#1+0.65*\dep,4+0.65*\dep) -- (4+#1+0.65*\dep,4+0.65*\dep);
        \draw[gray,thick,-] (4+#1+\dep,4+\dep) -- (4+#1+\dep,0+\dep);
        \draw[gray,thick,-] (4+#1+0.25*\dep,4+0.25*\dep) -- (4+#1+0.25*\dep,3+0.25*\dep);
        \draw[gray,thick,-] (4+#1+0.65*\dep,4+0.65*\dep) -- (4+#1+0.65*\dep,3+0.65*\dep);
}
\newcommand{\drawBlockBoundaries}[1]{
        \draw[green, dashed, very thick] (0+#1,0) rectangle (4+#1,4);
        \draw[green, dashed, very thick] (2+#1,0) -- (2+#1,4);
        \draw[green, dashed, very thick] (0+#1,2) -- (4+#1,2);

        \draw[green, dashed, very thick] (4+#1+1,0+1) -- (4+#1+1,4+1);
        \draw[green, dashed, very thick] (0+#1+1,4+1) -- (4+#1+1,4+1);

        \draw[green, dashed, very thick] (4+#1+2,0+2) -- (4+#1+2,4+2);
        \draw[green, dashed, very thick] (0+#1+2,4+2) -- (4+#1+2,4+2);

        \draw[green, dashed, very thick] (0+#1,4) -- (#1+\dep,4+\dep);
        \draw[green, dashed, very thick] (2+#1,4) -- (2+#1+\dep,4+\dep);
        \draw[green, dashed, very thick] (4+#1,4) -- (4+#1+\dep,4+\dep);

        \draw[green, dashed, very thick] (4+#1,0) -- (4+#1+\dep,0+\dep);
        \draw[green, dashed, very thick] (4+#1,2) -- (4+#1+\dep,2+\dep);
}

\journal{Computers \& Graphics}

\begin{document}

\verso{Preprint Submitted for review}

\begin{frontmatter}

\title{Direct interactive visualization of locally refined spline volumes\\for scalar and vector fields}

\author[1]{Franz G. \snm{Fuchs}\corref{cor1}}
\author[1]{Oliver J. D. \snm{Barrowclough}}
\author[1]{Jon M. \snm{Hjelmervik}}
\author[1]{Heidi E. I. \snm{Dahl}}
\cortext[cor1]{Corresponding author:
franzgeorgfuchs@gmail.com}
\address[1]{SINTEF Digital, Forskningsveien 1, 0314 Oslo, Norway}

\received{\today}

\begin{abstract}
    We present a novel approach enabling \emph{interactive} visualization of volumetric Locally Refined B-splines (LR-splines).
    To this end we propose a highly efficient algorithm for \emph{direct} visualization of scalar and vector fields given by an LR-spline.
    In both cases, our main contribution to achieve interactive frame rates is an acceleration structure for \emph{fast element look-up} and a change of basis for \emph{efficient evaluation}.
    To further improve the efficiency, we present a \emph{heuristic for adaptive sampling distance} for the numerical integration.
    A comparison with
    existing adaptive approaches is performed.
    The algorithms are designed to fully utilize modern graphics processing unit (GPU) capabilities.
    Important applications where LR-spline volumes emerge are given for instance by approximation of large-scale simulation and sensor data, and Isogeometric Analysis (IGA).
    We showcase interactive rendering achieved by our approach on different representative use cases, stemming from simulations of wind flow around a telescope, Magnetic Resonance (MR) imaging of a human brain, and simulations of a fluidized bed used for mixing and coating particles in industrial processes.
\end{abstract}


\end{frontmatter}


\setlength{\tabcolsep}{.12em}

\section{Introduction}
In recent years, there has been increasing industrial and scientific interest in splines, i.e., piecewise polynomial functions, in higher dimensions, driven by research efforts in managing \emph{large heterogeneous data sets} and advances in \emph{computational science}.
This has resulted in the development of several independent approaches to local refinement of splines in dimensions higher than one,
including hierarchical splines, 
T-splines, 
and LR-splines. 
The main advantage of local refinement is that it focuses approximation power locally where it is needed, allowing a considerable reduction in the amount of data needed to guarantee a given tolerance.

\emph{Large heterogeneous data sets} can originate from many different sources.
One example is \emph{geospatial data}, where existing and emerging data acquisition techniques provide a fast, efficient and affordable means for data collection.
The resulting raw data is generally very large and needs to be represented in a more suitable way e.g., for comparison and quality assurance.
Another example comes from \emph{simulations} in science, engineering and industry.
The design, validation and optimization of products and structures involves, e.g., physical, chemical, or biological processes that are modeled by partial differential equations (PDEs).
High-fidelity simulations are often necessary in order to perform a reliable analysis.
At the time of writing, the computational requirements and scalability of post-processing and visualization tools are a bottleneck in the simulation workflow.
The typical size of the simulation data is nowadays on the order of giga- or terabytes per time step, which is both impractical and inefficient to stream or download to a local client.
In both these two cases, locally refined splines have proven to be a good means to compactly represent these data, see, e.g., \cite{Skytt:2015}.
A quantitative and qualitative analysis of LR-spline approximations is the focus of an upcoming paper and will not be elaborated in this article.

In \emph{computational science} a recent development called isogeometric analysis (IGA), proposed by Cottrell, Hughes \& Bazilevs \cite{cottrell2009isogeometric}, has quickly become very popular in science and engineering.
IGA provides the integration of design and analysis by using a common representation -- splines -- for computer aided design (CAD) and finite element methods (FEM).
This eliminates the conversion step between CAD and FEM, which is estimated to take up to 80\% of the overall analysis time for complex designs \cite{cottrell2009isogeometric}.
In the context of IGA, local refinement of splines is an indispensable tool for efficient computations.

Despite the fact that locally refined splines have become very popular in a variety of situations, methods for \emph{visualization} of these types of splines are lagging behind or are nonexistent.
Direct volume and vector field visualization based on \locRefSpl{s} is completely novel.
In this paper we present a flexible framework for volumetric visualization based on volume and streamline rendering enabling interactive, direct visualization of volumetric LR-splines, that can be trivially extended to T-splines and hierarchical splines.
Our approach consists of several stages, leveraging the strengths of existing algorithms where possible.
Several features of our approach are novel:
\begin{enumerate}
    \item For fast evaluation of \locRefSpl{s} we extend the recently published method presented in \cite{Hjelmervik:2015} for interactive pixel-accurate rendering of LR-spline surfaces to the volumetric case.
        The extension has different issues compared to the two dimensional case.

    \item We compare and discuss the pros and cons of different acceleration structures for LR-spline element look-up: textures, octrees, k-d trees, and k-d forests. 
        The details of LR-splines in 3D vary sufficiently over the 2D case to warrant its own treatment here.
    \item We present a heuristic for adaptive sampling distance, based directly on the LR-spline structure, and compare it with existing approaches.
\end{enumerate}

The rest of the paper is organized as follows. Related work is discussed in \refsec{relatedwork}.
An algorithm for efficient evaluation of LR-splines is described in \refsec{lrsplines}.
Applications are presented and details of the performance of the implementation of the overall algorithm for interactive volume rendering are described in \refsec{volrender} and for vector field visualization in \refsec{vecrender}.
Finally, we draw conclusions based on the results in \refsec{conclusion}.

\section{Related work}\label{sec:relatedwork}
Scientific volume visualization techniques convey information about a vector or scalar field defined on a given geometry.
The techniques can be divided into the following approaches: iso-surface extraction and volume rendering for \emph{scalar fields}; direct approaches, e.g. glyphs, line integral convolution (LIC), and stream-/streak-/pathline rendering for \emph{vector fields}.
In this article we focus on volume and stream-/streak-/pathline rendering.

The main challenge for achieving interactive volume visualization in the setting of \locRefSpl{s} is that sampling is computationally expensive due to the need for spline evaluation.
However, they allow for a compact representation, particularly when local refinement is applied.

\subsection{Scalar field visualization}

The visualization of a scalar field is commonly done by modeling the scalar field as a participating medium, where a modifiable transfer function specifies how field values are mapped to emitted color and transparency.
In the simplest case, where the field consists of discrete samples over a regular grid, an abundance of results is available, see e.g., \EC{Levoy~}\cite{Levoy:1988:DOS} for an early example or \EC{Engel at al.~}\cite{engel2006real} for an overview.
Octrees allow a voxel grid to have different resolutions across the domain, which reduces the amount of data needed for a given scene.
This is used in e.g., GigaVoxels \cite{crassin2009gigavoxels}, which offers real-time rendering of several billion voxels.
GigaVoxels is only effective when there is an a priorily known location of significant regions of empty space.

Kreylos et al. focus on direct volume rendering of adaptive mesh refinement (AMR) data, but the types of mesh are heavily restricted \cite{kreylos2002remote}.
High quality rendering of more general AMR data is presented by Marchesin \& De Verdiere \cite{marchesin2009high}; the authors report 4 fps on screen resolutions up to $2048^2$ on a data set with 2377878 cells with 4 levels of refinement.
For a recent survey on state-of-the-art GPU-based large-scale volume visualization we refer to \cite{beyer2014survey}.

\begin{figure*}
    \centering
    \includegraphics[height=.18\textheight]{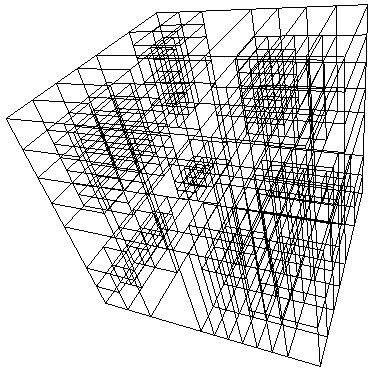}
    \includegraphics[height=.18\textheight]{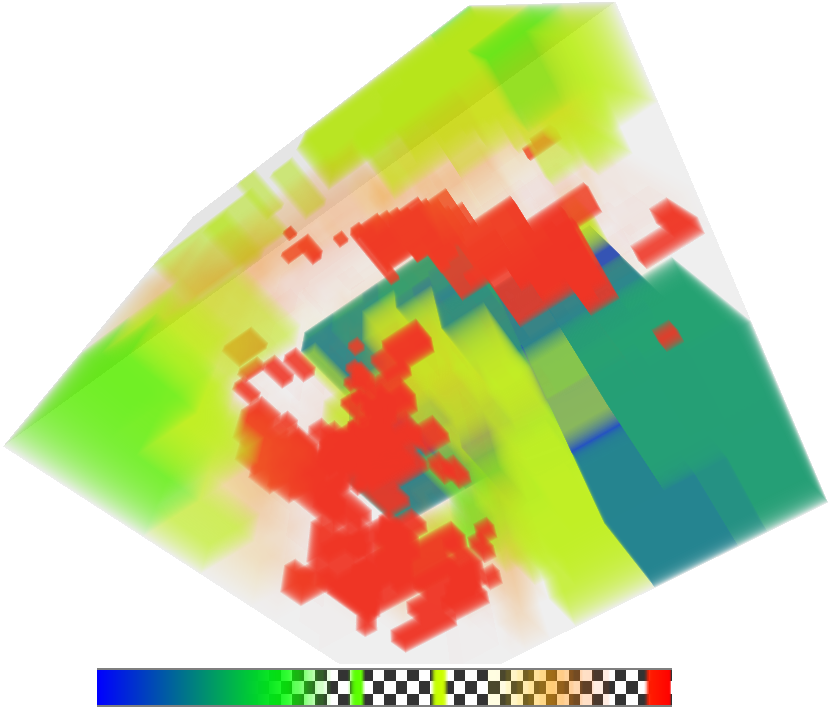}
    \caption{
        Visualization of local box size with very coarse (left) and medium refinement (right); blue indicates large and red small box size.
    }
    \label{fig:boxsize}
\end{figure*}

\subsection{Vector field visualization}
In the case of vector fields, occlusion and complexity make direct visualization of the whole data set very hard to interpret visually.
Interactive texture based flow visualization was one of the first use cases for programmable graphics hardware~\cite{Weiskopf01hardware-acceleratedvisualization}.
Such methods take advantage of and rely on fast texture lookups and regular memory operations to obtain good performance, which limits the use to uniform grids.

For glyphs and stream-/streak-/pathline rendering, seeds are typically chosen directly by the user, or are based on simplification or feature extraction.
The main requirement for interactive implementation of these techniques is fast evaluation of the vector field.
In this paper we will focus on the evaluation algorithm of volumetric LR-spline vector fields and apply it to streamline rendering.
However, the LR-spline approach is applicable to other visualization methods as well.

\section{LR-splines}\label{sec:lrsplines}

In the context of this paper, LR-splines provide a compact representation for modelling generally smooth functions with detail that varies over multiple scales, since they support both local refinement and higher degree representations.
LR-splines are a superset of the commonly used tensor-product splines and in fact, any LR-spline is generated by making local refinements of a tensor-product spline mesh.
For a detailed overview of the general theory of LR-splines see \cite{Dokken:2013,Johannessen:2014}.

\subsection{Background}
 
Spline functions $\mathbf{s}:\Omega\subset\mathbb{R}^m\rightarrow\mathbb{R}^n$ are piecewise polynomials that can be defined in any dimension $m,$ and over a wide variety of spatial partitions (e.g., box or simplex partitions).
Univariate splines ($m=1$) of a given degree $p,$ are very often formulated in terms of B-spline basis functions $(B_{U,i}^p(x))_{i=1}^{l},$ which are in turn defined from a non-decreasing sequence of real values, $U=\{u_1,\ldots,u_{l+p+1}\},$ known as a knot vector.
The purpose of the knot vector is to partition the domain interval $\Omega=[u_1,u_{l+p+1}]$ into segments on which the spline function is polynomial and to determine the degree of continuity of the function between polynomial segments.
Univariate splines have a straightforward generalization to multivariate tensor-product splines, 
where each dimension can be represented by a univariate spline.
As an example, for $m=3,$ a vector-valued spline $\mathbf{s}$ of tri-degree $\mathbf{p}=(p_1,p_2,p_3)$ can be expressed in terms of tensor-product B-splines as
\begin{equation}\label{eq:Bspline}
    \mathbf{s}(x,y,z) = \sum_{i=1}^{l_1} \sum_{j=1}^{l_2} \sum_{k=1}^{l_3} \mathbf{c}_{i,j,k} B_{U,i}^{p_1}(x)B_{V,j}^{p_2}(y)B_{W,k}^{p_3}(z),
\end{equation}
where $\mathbf{c}_{i,j,k} \in \R^{n}$ are the control points, or coefficients, and the knot vectors $U=\{u_1,...,u_{l_1+p_1+1}\}$, $V=\{v_1,...,v_{l_2+p_2+1}\}$ and $W=\{w_1,...,w_{l_3+p_3+1}\}$ are given.
For a more detailed introduction to splines see \cite{Farin:2002}.

Despite its simplicity, the tensor-product form has some downsides when it comes to refinement.
If a new knot value is inserted in one of the directions, the effect pervades throughout the entire domain in all other directions.
The result is a global increase of the size of the representation, even when only local modelling flexibility is needed.
This limits the value of tensor-product splines to relatively small examples, with few knots and low spatial dimension $m$ of the domain.

The increase of interest in splines in higher dimensions, driven by research efforts in fields such as isogeometric analysis, has resulted in several independent approaches to adaptive refinement of spline spaces in dimensions $m < 1$.
These include hierarchical splines originally developed in \cite{Forsey:1988} and extended in \cite{Giannelli:2012}; T-splines developed in \cite{Sederberg:2003}; and LR-splines developed in \cite{Dokken:2013}.
Though the details of the various methods differ, they all offer the ability to add degrees of freedom to the spline space locally in order to increase modelling flexibility.
The pre-processing step in this paper is based on LR-splines, though the methods can easily be applied to both T-spline and hierarchical spline models.

A vector-valued volumetric LR-spline function, $\mathbf{f}:\Omega\subset\R^{3} \rightarrow \R^{n}$ of tri-degree $\mathbf{p}=(p_1,p_2,p_3)$ is defined as a linear combination of functions known as LR B-splines, $N^\mathbf{p}_i$:
\begin{equation}\label{eq:lrspline}
    \mathbf{f}(x,y,z) = \sum_{i\in\Xi} \mathbf{c}_i \gamma_i N^{\mathbf{p}}_i(x,y,z),
\end{equation}
where $\gamma_i$ are scaling factors that ensure a partition of unity and $\mathbf{c}_i \in \R^{n}$ are the coefficients.
Rather than defining knot vectors in each dimension, LR B-splines are inferred from an LR-mesh, as pictured in \reffig{boxsize} (left).
Each LR B-spline is locally equivalent to a tensor-product B-spline, and they are generated by recursively splitting every LR B-spline on the mesh until all LR B-splines are minimal support.
Since the set of LR B-splines $N^{\mathbf{p}}_i$ has no unique natural ordering, we label them by an index set $\Xi$.
The domain $\Omega$ of $\mathbf{f}$ in $\R^3$ is a \emph{box}, i.e., a product of three one-dimensional intervals, which we can assume to be $[0,1]^3$. 
The domain is partitioned into a set of three-dimensional axis parallel boxes or \emph{elements} known as a box partition. 
An example of a volumetric LR-spline box partition is pictured in \reffig{boxsize} (right).

LR-splines have similar properties to tensor-product splines, such as minimal support basis functions and partition of unity.
However, linear independence of the set of LR B-splines is not guaranteed and is a challenging problem for general refinements in three dimensions \cite{Pettersen:2013}.
Linear independence is critical for applications in IGA, but for the purposes of visualization it does not pose a problem.

\begin{figure*}
    \centering
    \begin{tikzpicture}[sibling distance=25pt,scale=0.7]
        \tikzset{level distance=20pt}
        \drawLRelements{0}
        \node (a) at (2.5,1.5) {$e_0$};
        \node (a) at (0.5,2) {$e_1$};
        \node (a) at (1.5,3.5) {$e_2$};
        \node (a) at (3,3.25) {$e_3$};
        \node (a) at (2.5,3.75) {$e_4$};
        \node (a) at (3.5,3.75) {$e_5$};
        \node (a) at (2.2,5.30) {$e_6$};
        \node (a) at (3.8,5.30) {$e_7$};
        \node (a) at (4.3,4.85) {$e_8$};
        \node (a) at (5.1,5.6) {$e_9$};
        \draw[->, >=latex, blue!20!white, line width=5pt] (5.2,1) to (6.8,1);
        \def\picdist{7}
        \drawLRelements{\picdist}
        \drawBlockBoundaries{\picdist}
        \node (element00) at (\picdist+1,1) {\color{green} $P_{000}$};
        \node (element00) at (\picdist+3,1) {\color{green} $P_{010}$};
        \node (element00) at (\picdist+1,3) {\color{green} $P_{001}$};
        \node (element00) at (\picdist+3,3) {\color{green} $P_{011}$};

        \node (element00) at (\picdist+2.5,3+2.5) {\color{green} $P_{101}$};
        \node (element00) at (\picdist+4.5,3+2.5) {\color{green} $P_{111}$};
        \node (element00) at (\picdist+5.5,2+0.5) {\color{green} $P_{110}$};
        \draw[->, >=latex, blue!20!white, line width=5pt] (12.2,1) to (13.8,1);
        \node (element000) at ({2*\picdist+1.5},6) {\color{green} $P_{000}$}
            child {node {$x<0.25$}
                child{ node{$e_1$}}
                child{ node{$e_0$}}
            };
        \node (elementxxx) at ({2*\picdist+1.5},4) {.};
        \node (elementxxx) at ({2*\picdist+1.5},3.9) {.};
        \node (elementxxx) at ({2*\picdist+1.5},3.8) {.};
        \node (element111) at ({2*\picdist+1.5},3.2) {\color{green} $P_{111}$}
            child {node {$x<0.75$}
                child{ node{$z<0.75$}
                    child[opacity=0]{ }
                    child{ node{$e_0$}}
                    child{ node{$e_7$}}
                    child[opacity=0]{ }
                    child[opacity=0]{ }
                }
                child[opacity=0]{ }
                child{ node{$y<0.65$}
                    child[opacity=0]{ }
                    child{ node{$z<0.75$}
                        child{ node{$e_0$}}
                        child{ node{$e_8$}}
                    }
                    child[opacity=0]{ }
                    child{ node{$z<0.75$}
                        child{ node{$e_0$}}
                        child{ node{$e_9$}}
                    }
                }
            };
    \end{tikzpicture}
    \caption{Fast element search for \locRefSpl{s}. Left: Elements $e_l$. Middle: Regular blocks $P_{i,j,k}$. Right: K-d Search forest.}
    \label{fig:kdforest}
\end{figure*}
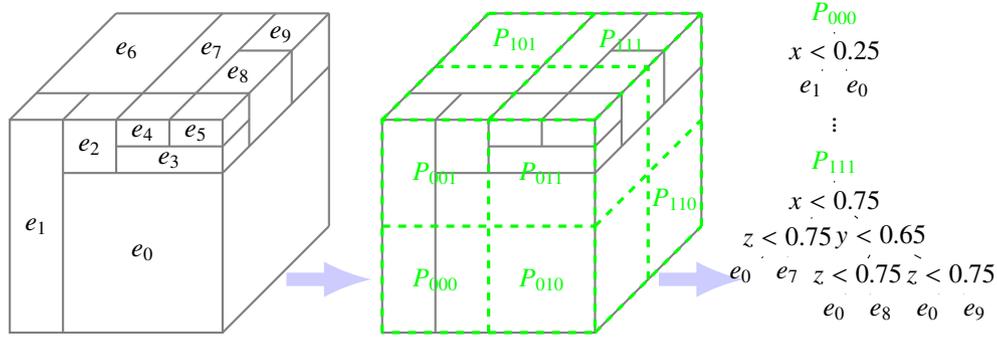

\subsection{Fast LR-spline evaluation on the GPU}\label{sec:fasteval}
\locRefSpl{s} provide a flexible representation for compactly modelling volumetric data with overall smooth features and high local variation.
However, a drawback of this flexibility is that the evaluation becomes more involved, when compared to tensor-product splines.
This is due to the irregular nature of the underlying LR-mesh. 
To remedy this, we extend the approach presented in \cite{Hjelmervik:2015} to three dimensions.

Given a point in the parameter domain, one needs to first find the corresponding element $e_l$, see illustration in \reffig{kdforest}, (left).
For a large number $N$ of elements, a linear search through all elements, with its complexity of $\mathcal{O}(N)$ in the average case, quickly becomes infeasible in the context of real-time rendering.
Let the parameter domain be $[0,1]^3$ for the following discussion.
For quick element look-up
\begin{itemize}
    \item \textbf{Texture.}
        For the special case where all knots in each direction can be written as multiples of $1/l^i$ where $l,i$ are natural numbers -- typically $1/2^i$ -- a look-up texture is the fastest method with a complexity of $\mathcal{O}(1)$.
        However, it is apparent that this approach fails in many cases, e.g., if a knot value is an irrational number.
        In addition, the maximum size of OpenGL textures imposes a strict limit on the levels of local refinement.
    \item \textbf{Octree.}
        An octree can be used to remedy the restriction on the maximum level of refinement of the texture based approach.
        The search complexity of octrees is $\mathcal{O}(\log(N))$.
        The downside of octrees with respect to LR-spline meshes is that they can only represent a regular data structure.
        In fact octrees only work if all knots can be written as multiples of $1/2^i$.
    \item \textbf{K-d Tree.}
        K-d trees are an excellent data structure for axis aligned splines with no restrictions on the knot values, i.e., for a general LR-spline.
        The average search complexity is $\mathcal{O}(\log(N))$.
        However, since knot values can be at arbitrary parameter points, the split value needs to be stored and read from memory as well.
    \item \textbf{K-d Forest.}
        A k-d forest consists of a regular block division of the domain, where each block has its own k-d tree, see \reffig{kdforest} (right).
        This approach works for general LR-splines at arbitrary refinement levels (just as one k-d tree), but also allows exploitation of the efficiency of texture look-ups whenever possible.
        More details are presented in \refsec{kdforest}.
\end{itemize}
For all cases, the space requirement for the structure is $\mathcal{O}(N)$.
The chosen approach consists of a k-d forest acceleration structure to identify the element containing a given parameter value, and a \Bezier representation of the field in each element for efficient evaluation.
A comparison of the different approaches is presented \refsec{volresults} and \refsec{vecresults}.
These two components are described in the following sections.

\subsubsection{K-d forest}\label{sec:kdforest}
In order to handle element look-up for general LR-splines efficiently, we present an approach that combines the speed of texture look-ups, with the necessary flexibility to handle general LR-splines provided by k-d trees.
To this end, the parameter domain is divided into a set of $I\times J\times K$ regular blocks $\{P_{i,j,k}\}$.
For each of these blocks we build a k-d tree for all elements that intersect the block.
\reffig{kdforest} schematically depicts the acceleration structure of the resulting k-d forest, i.e., multiple \emph{k-d trees}.
The influence of the number of trees on the evaluation time is discussed in \refsec{volrender} and \refsec{vecrender}.

\begin{algorithm}[t]
    tmp = access texture at value $p$\;
    isLeaf = (32st bit of tmp == 1)\;
    \eIf{isLeaf}
    {
        elementNumber = first 29 bits of tmp\;
        done\;
    }
    {
        nextIndex = first 29 bits of tmp\;
    }
    \While{true}{
        kdnode = texelFetch(forest, nextIndex).rg\;
        isLeaf = (32st bit bit of kdnode.r == 1)\;
        \eIf{isLeaf}
        {
            elementNumber = first 29 bits of kdnode.r\;
            done\;
        }
        {
            dir ($x, y, or z$) = bits 30 and 31 of kdnode.r\;
            \eIf{ $p_{dir} < $kdnode.g }
            {
                nextIndex = first 29 bits of kdnode.r;
            }
            {
                nextIndex += 1;
            }
        }
    }

    \caption{Algorithm (k-d forest) to obtain the element number, given a parameter value p.}
    \label{alg:fasteval}
\end{algorithm}

The k-d forest is precomputed on the CPU and uploaded to the GPU.
The head of each search tree is stored in a 3D texture for rapid look-ups in the search forest.
This texture has the size $I\times J\times K$ with internal format GL\_R32UI.
In the case of a single active element in a block $P_{i,j,k}$ the element index is stored directly, otherwise it is interpreted as a pointer to the start of the search tree.
The search forest is uploaded to a GL\_RG32F texture buffer.
The red component is used to store the indices/leafs, the green to store the split value.
To minimize the amount of storage needed, we use the high bits to store if the node is a leaf or not, and if the split is in the x-, y-, or z-direction.
A pseudo-algorithm for identifying the element containing a parameter value $p$ is given in \refalg{fasteval}.

In the worst case, i.e., for unbalanced trees, the search complexity is increased to $\mathcal{O}(N)$.
When creating the search forest, it is therefore important to create \emph{depth-balanced} trees, in order to ensure the search complexity of $\mathcal{O}(\log(N))$.
In order to achieve this, the correct direction to split, i.e., x-, y-, or z-direction, must be found.
We employ an algorithm that is based on counting the active elements to the left and right of the mean split value in each direction.

\subsubsection{Octree}
In the special case, when all knots are can be written as multiples of $1/2^i$ (assuming a parameter domain $[0,1]^3$), an octree can be used for efficient element look-up.
The octree is precomputed on the CPU and uploaded to the GPU as a GL\_R32F texture buffer.
The generation of the octree is straightforward;
the algorithm starts with the whole parameter domain,
 then subdivides the regions successively into 8 octree regions, whenever a region overlaps with more than one LR element.
To minimize the amount of storage needed, we use a similar approach to the one presented for k-d tree/forest;
the highest bit of each entry in the octree data structure can be used to indicate if the entry is a leaf or an element number.

\subsubsection{\Bezierblock{s}}
In order to improve the efficiency of evaluating LR-splines, we convert to a representation that is locally more regular, known as \Bezierblock{s}. 
Since the transformation from LR B-splines to \Bezierblock{s} is a change of basis rather than an approximation, the representation quality is maintained, as well as the continuities between adjacent blocks.
This basis shift can easily be done by computing the coefficients of an interpolating polynomial.
In addition to the coefficients, the value of the lower left and upper right corner of each element has to be stored on the GPU memory in a texture buffer.
There are 4 main reasons for this change of representation:
\begin{itemize}
    \item LR-splines are based on irregular data structures, with coefficients distributed in memory. This is not optimal for an implementation on the GPU.
    \item The number of basis functions is the same for all blocks when evaluating in \Bezier form, which is well-suited for evaluation on the GPU.
        In general, the number of LR B-spline basis functions (i.e., the local dimension) can vary from element to element and is greater or equal to the number of basis functions needed to evaluate the same spline in \Bezier form.
    \item LR-splines cannot be evaluated in tensor-product form: for a spline of tri-degree $\mathbf{p}$, evaluating the same \block in \Bezier form needs $3+p_1+p_2+p_3$ one-dimensional basis evaluations, while an LR-spline needs at least $(p_1+1)(p_2+1)(p_3+1)$.
    \item An added advantage of this conversion is that the algorithm and its implementation becomes independent of the representation; LR-splines, T-Splines, and hierarchical splines can all be converted to \Bezierblock{s} and our algorithm applied for rendering.
\end{itemize}
With this strategy, well known algorithms, such as De Casteljau's algorithm, can readily be used to evaluate the volumetric splines on the GPU.

\section{Volume rendering}\label{sec:volrender}
Volume rendering is based on tracing view-rays through the volume from an imaginary observer.
If such a view-ray intersects the object one obtains the color for the pixel of the screen by evaluating an integral describing the accumulated radiance along the ray.
When taking both emission and absorption into account, the accumulated radiance $I_\lambda$ for wavelength $\lambda$ along a view-ray $\gamma:\R\rightarrow\R^3$ is given by the so-called volume render integral
\begin{equation}
    \label{eq:volRenderIntegral}
    I_{\lambda}(t) = I_{\lambda}(0) T_{\lambda}(0,t) + \int_{ \left.\gamma\right|_{[0,t]} } \!\!\! \!\!\! \glow_{\lambda}(s) T_{\lambda}(s,t) ds,
\end{equation}
where $\int_\gamma$ denotes the line integral.
The function $\glow_\lambda(s)$ specifies emission, and $T_\lambda(s,t)$ specifies absorption (from $s$ to $t$) of light with the wavelength $\lambda$.
In applications, one typically uses three groups of wavelengths representing red, green, and blue.
Emission and absorption, defined by a so-called transfer function, depend on the value of the scalar field $\rho$.
Here, we extend classic volume rendering to \locRefSpl{s}.

\subsection{Background: discretizations of the volume render integral}
Discretizations of \refequ{volRenderIntegral} are based on splitting the integral into intervals.
Efficient implementations on GPUs are so-called compositing schemes where color and opacity is accumulated iteratively.
Front-to-back compositing
for the accumulated radiance $C_{dst} = (I_r,I_g,I_b)^T$, and the accumulated opacity $\alpha_{dst} = (1-T_{dst})$
is given by \refalg{compositing}.

\begin{algorithm}
    \caption{Front-to-back compositing}
    \label{alg:compositing}
\begin{algorithmic}
    \State $T \gets (1-\alpha_{src})^{\Delta s_i/\xi}$
    \State $C_{dst} \gets C_{dst} + (1-T) (1-\alpha_{dst}) C_{src}$
    \State $\alpha_{dst} \gets\, \alpha_{dst} + (1-T) (1-\alpha_{dst})$
\end{algorithmic}
\end{algorithm}

\noindent
Here, $\Delta s_i$ is the varying ray segment length and $\xi$ is a standard length.
Furthermore, $C_{src}$ and $\alpha_{src}$ are given by the transfer function.

In many application areas transfer functions contain high frequency components, dictating a high sampling rate (Nyquist rate).
One method is \textit{oversampling}, i.e., introducing additional sampling points, although the underlying scalar field is approximately linear.
Evaluating the scalar function in the setting of locally refined spline volume rendering is a time-intensive operation as it means evaluating \refequ{lrspline}.
To avoid oversampling, a common technique is \textit{pre-integration} (see e.g., \cite{roettger2003smart,Engel:2001}), which is based on calculating the volume render integral for pairs of sample values in advance.
In this article we employ a technique called \textit{supersampling} presented in\cite{Fuchs:2015}, in order to account for the non-linearity of $I$.
The approach is easy to implement, and successfully captures high frequencies of the transfer function.
This increases the image quality by reducing so-called wood-grain artifacts, while avoiding computationally expensive evaluations of the \locRefSpl function.

An important tool for investigating a data set is to hide insignificant regions.
This is particularly important when LR-spline volumes, that are axis-aligned, come from approximating simulation data on arbitrary geometries.
The values outside the valid geometry are artificial and need to be hidden.
We trim the rectilinear geometry of the LR-spline volume by the boundary mesh of the original data.
For a tetrahedral simulation mesh, this boundary mesh is a triangulated surface which can be represented by an STL object.
To achieve the \emph{trimming}, we employ a two-stage algorithm, known as \emph{z-prepass}.

\begin{figure}
    \centering
    \subfigure[Standard Volume Rendering.]{
        \includegraphics[width=0.45\linewidth]{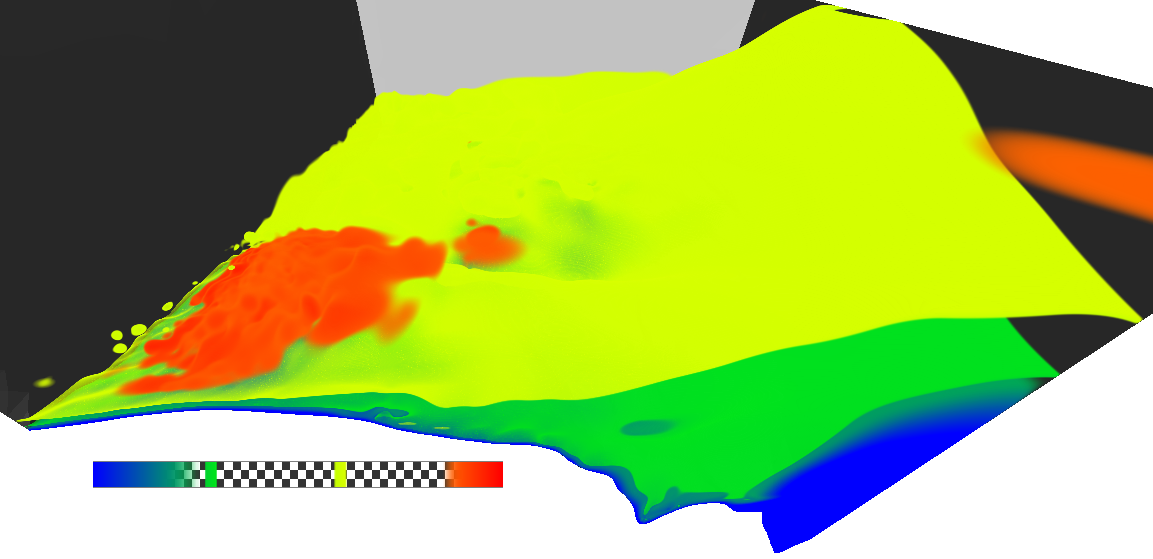}
    }
    \subfigure[Local illumination.]{
        \includegraphics[width=0.45\linewidth]{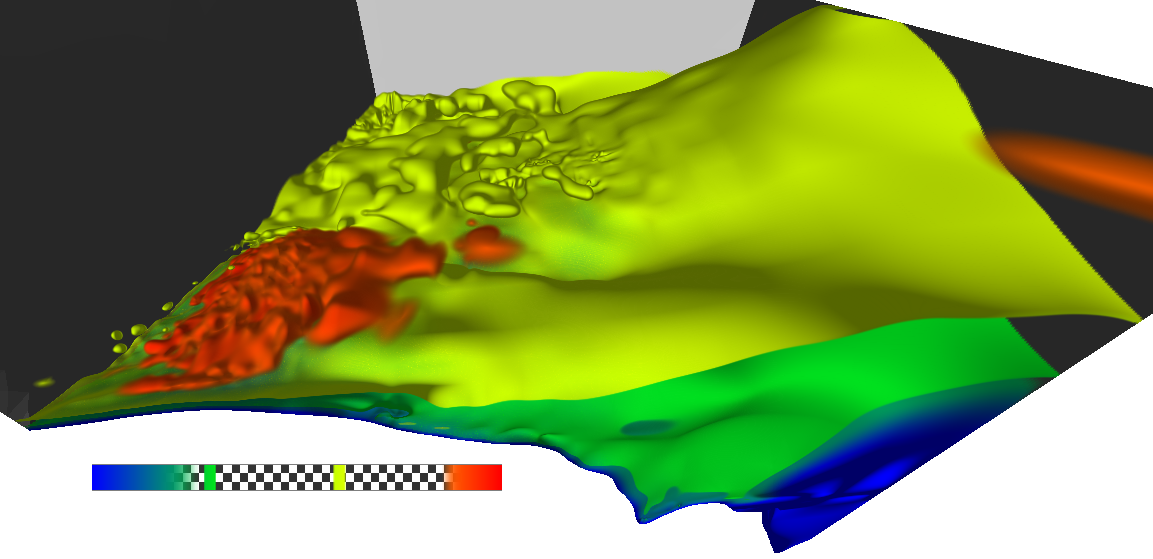}
    }
    \caption{Local light illumination helps to reveal important geometric information of the scalar field. The normal direction is readily available for higher order splines.}
    \label{fig:lli}
\end{figure}

In order to enhance the perception of the data, \emph{gradient-based local illumination} can be used.
This lighting model assumes that external light is reflected at isosurfaces inside the volumetric data.
Note that at each sample point both the scalar value and the gradient of the volume has to be evaluated.
For an LR-spline or \Bezier block, the \emph{gradient can be evaluated exactly and cheaply} by reusing calculations for the scalar value.
Applying this model to \refalg{compositing} results in multiplying $C_{src}$ in step two with the local light intensity $I(x,y,z)$.
We present results using the diffusive lighting model given by $I(x,y,z) = \max(0.4, |\vec{l}\cdot \vec{n}(x,y,z)|))$, where $\vec{l}$ is the vector to the light source and $\vec{n}(x,y,z)$ is the (local) normal direction.
An example to show how much more geometric information can be revealed with local illumination is presented in \reffig{lli}.

\begin{figure}
    \centering
    \subfigure[
        The bars show the colors that are assigned to the values of the scalar field, where a checkerboard pattern indicates transparent regions.
    ]{
        \includegraphics[width=1.0\linewidth]{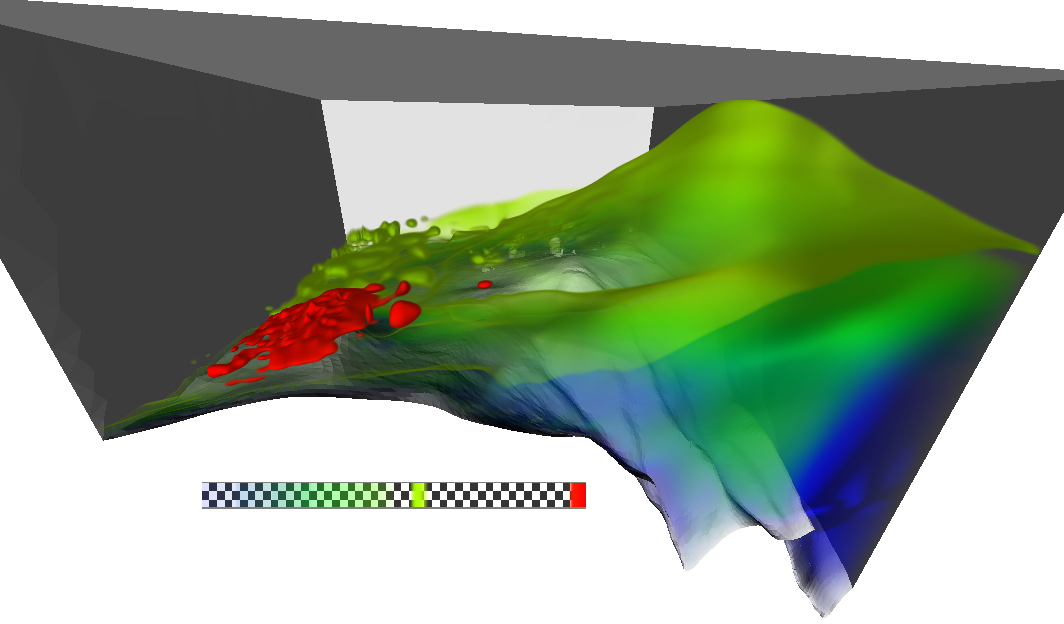}
    }
    \subfigure[
        Computation times for LR-spline evaluation for a representative approximation comparing different search methods.
        The computation was performed on an NVIDIA Titan GPU and a screen resolution of $1280\times720$.
        ]{
        \begin{tabular}{| l || c || c || c || c | c | c |}
            \hline
            \multicolumn{1}{|c||}{element search}& texture & octree & \multicolumn{1}{|c||}{k-d tree} & \multicolumn{3}{|c|}{k-d forest}  \\
            \hline
            \# trees$^3$     &      &      &  1 &  8   & 64    &  512  \\
            min depth        &      &      &    & 0    &  0    &  0    \\
            max depth /depth &      &   12 & 83 & 43   & 30    & 6     \\
            mean depth       &      &      &    & 5.3  & 0.07  & 9.6e-5\\
            var depth        &      &      &    &  5.2 &  0.5  &  0.02 \\
            \hline
            evaluation [ms]   &  -  &  99   &100 & 55   & 30    & 25    \\
            \hline
        \end{tabular}
    }
    \caption{
        Volume Rendering of wind speed around telescope.
    }
    \protect\label{fig:lrvol1}
\end{figure}

\begin{figure}
    \subfigure[
        The bars show the colors that are assigned to the values of the scalar field, where a checkerboard pattern indicates transparent regions.
    ]{
        \includegraphics[width=1.0\linewidth]{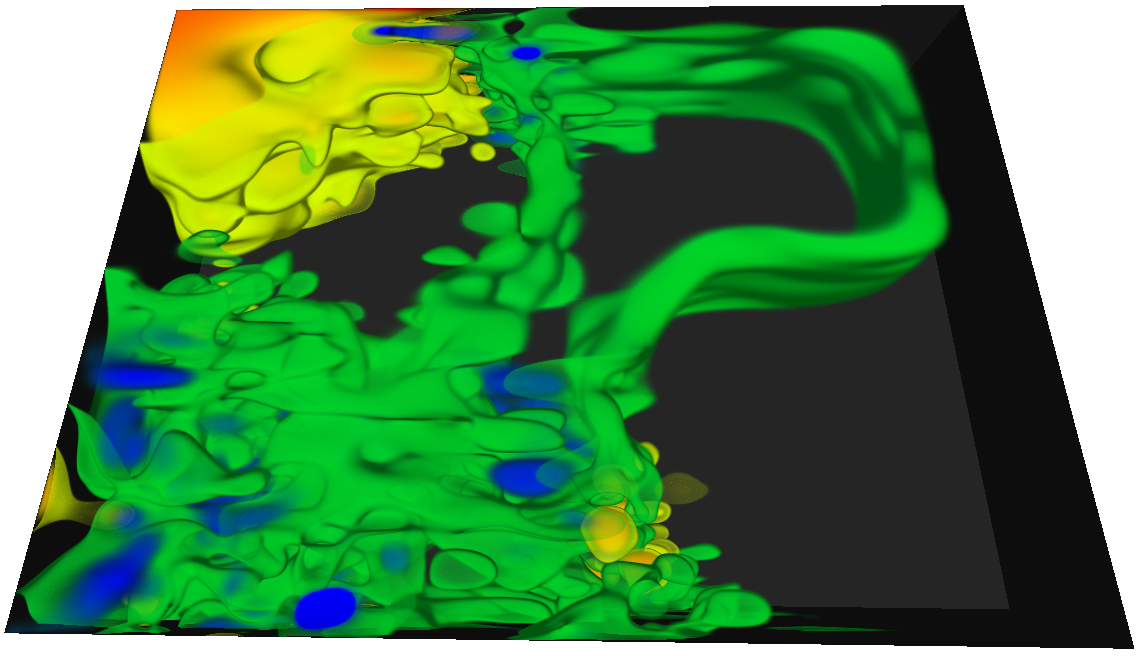}
    }
    \subfigure[
        Computation times for LR-spline evaluation for a representative approximation comparing different search methods.
        The computation was performed on an NVIDIA Titan GPU and a screen resolution of $1280\times720$.
        ]{
        \begin{tabular}{| l || c || c || c || c | c | c |}
            \hline
            \multicolumn{1}{|c||}{element search}& texture & octree & \multicolumn{1}{|c||}{k-d tree} & \multicolumn{3}{|c|}{k-d forest}  \\
            \hline
            \# trees$^3$       & & &  1 &  8 & 64 & 512 \\
            min depth          & & & 34 &  0 & 0 & 0  \\
            max depth  / depth & & & 34 & 10 & 4 & 4 \\
            mean depth         & & & 34 & 5.2 & 0.9 & 0.1 \\
            variance depth     & & & 0 & 2.1 & 1.1  & 0.5 \\
            \hline
            evaluation [ms] (no illum.) & - & - & 77 & 34 & 24 & 28 \\
            \hline
        \end{tabular}
    }
    \caption{
        Volume Rendering of the particle speed in a fluidized bed.
    }
    \protect\label{fig:lrvol2}
\end{figure}
\begin{figure}
    \subfigure[
        The bars show the colors that are assigned to the values of the scalar field, where a checkerboard pattern indicates transparent regions.
    ]{
        \includegraphics[width=1.0\linewidth]{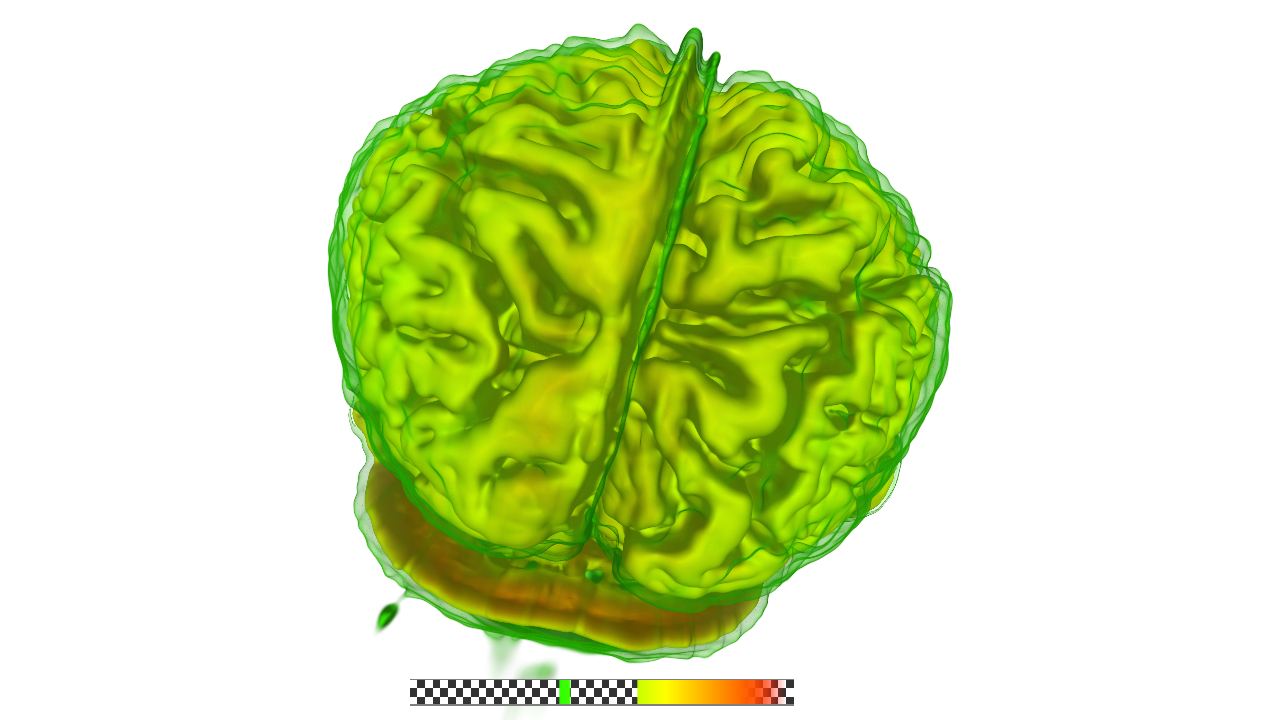}
    }
    \subfigure[
        Computation times for LR-spline evaluation for a representative approximation comparing different search methods.
        The computation was performed on an NVIDIA Titan GPU and a screen resolution of $1280\times720$.
        ]{
        \begin{tabular}{| l || c || c || c || c | c | c |}
            \hline
            \multicolumn{1}{|c||}{element search}& texture & octree & \multicolumn{1}{|c||}{k-d tree} & \multicolumn{3}{|c|}{k-d forest}  \\
            \hline
            \# trees$^3$ / texture size$^3$     & 128&      &  1 &  4   &  16 & 64    \\
            min depth                           &    &      &    &  6   &   0 &  0    \\
            max depth / depth                   &    & 8    & 89 & 53   & 18  &  3    \\
            mean depth                          &    &      &    & 28.0 & 6.4 &  0.75 \\
            var depth                           &    &      &    & 14.0 & 6.0 &  1.2  \\
            \hline
            evaluation [ms]                      & 38 & 279  & 188 & 105 & 58 & 42   \\
            \hline
        \end{tabular}
    }
    \caption{
        Volume rendering of an MR-brain scan.
    }
    \label{fig:lrvol3}
\end{figure}

\subsection{Heuristic for adaptive sampling distance}\label{sec:adaptivesd}
For LR-splines the size of the elements, e.g., as measured by diagonal length, can vary by several orders of magnitude.
In the instance of the wind simulation presented in \reffig{lrvol1} the ratio between the largest and the smallest diagonal is 2048.
A uniform sampling distance $\Delta s_i$ dictated by the smallest elements quickly becomes prohibitive for real time rendering.
Because of the local refinement strategies of LR-splines, the highest level of refinements, i.e., the smallest elements, occur in areas with the largest local variation in the data.
Thus the degree of variation of the field within each element can be expected to be of the same order of magnitude.
As a result, an adaptive sampling $\Delta s_i$ in \refalg{compositing} necessary for "capturing" all local details, is proportional to the local element size.
We obtain an efficient algorithm by choosing the local step size to be a function of the length of the box sides at the point $p\in\R^3$, i.e.,
\begin{equation}
    \label{eq:localss}
    \Delta s_i (p) = \frac{1}{2^d} \underset{\alpha\in\{x,y,z\}}{\min}\{\text{boxside}(p)_\alpha)\},
\end{equation}
where $d$ is the degree of the spline in this element.

\subsection{Results}\label{sec:volresults}
In this section we provide several examples of data sets given by LR-spline volumes.
We analyze the rendering performance of the different types of acceleration structure, described in \refsec{lrsplines}, and different methods for adaptive and non-adaptive sampling described in this section.

A screenshot for the case of the \emph{wind speed simulation around a telescope} is presented in \reffig{lrvol1}(a).
The LR-volume has tri-degree (2,2,2) and consists of 36614 elements with 256 levels of refinement, i.e., the smallest box size is 256 times less than that of the largest.
In this case the LR-volume needs to be trimmed by a geometry, given by an STL file.
This trimming, i.e., the rendering of the front and back faces, takes about 2.2 ms for an STL comprising 2.4 million primitives.

\reffig{lrvol1}(a) shows a comparison of rendering times with respect to the different element look-up strategies.
Look-up textures are not applicable in this case, since the level of refinement is too high.
Octree and k-d tree structures achieve approximately the same rendering time, where the octree has less depth but more children.
K-d forest structure is the fastest element-search method.
The table shows that the minimum depth, as well as the mean and variance of the trees quickly approaches zero with increasing number of trees in the k-d forest.
This results in a decrease in the rendering times, as there are more and more cases where the element search can be performed by a single texture look-up instead of tree-traversal.
However, even a texture size of $512^3$ is not sufficient to avoid the usage of k-d trees.

\reffig{lrvol2} shows a screen shot for the case of \emph{simulation of a fluidized bed}.
Particle velocity is given by an LR-spline of tri-degree (2,2,2) with 5311 elements and 13 refinement levels.
In this case, neither textures, nor octrees are applicable, since there are knot values that cannot be written as multiples of $1/l^i$ where $l,i\in \N$.
The rendering time improves with increasing number of trees in the forest, see \reffig{lrvol2}(b).
In addition, it can be deduced that local volume illumination increases the rendering time by approximately 80\%, but allows to reveal important geometric information.

Finally, we present an \emph{MR brain scan} given by an LR spline of tri-degree (2,2,2) with 255909 elements and 16 levels of refinement.
\reffig{lrvol3} shows the rendering times for the different element-search algorithms.
A look-up texture is the most efficient acceleration structure, but a k-d forest with $64^3$ trees is almost equally fast; the octree algorithm is the slowest.

\begin{figure}
    \centering
    \subfigure[
        Visualization of particles in a fluidized bed, with color indicating speed from low (blue) to high (red).
        The rendering stage of the streamlines on a screen with resolution of $1280\times720$ takes about 7ms.
        ]
    {
        \includegraphics[width=.8\linewidth]{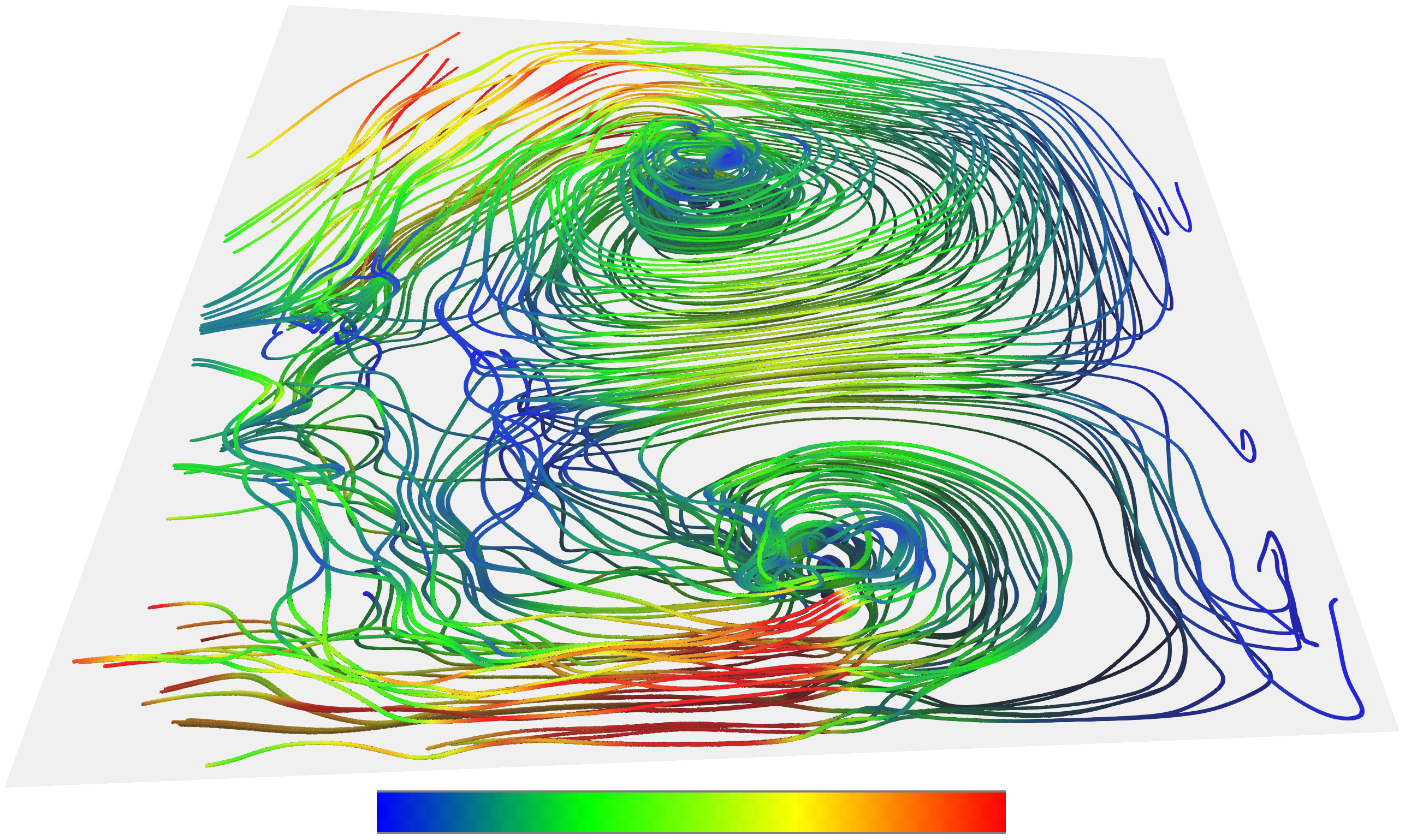}
    }
    \subfigure[
        Computation times for LR-spline evaluation for a representative approximation comparing different search methods.
        The measurement was performed on an NVIDIA Titan GPU.
        ]
    {
    \begin{tabular}{| l || c || c || c || c | c | c |}
        \hline
        \multicolumn{1}{|c||}{}& texture & octree & \multicolumn{1}{|c||}{k-d tree} & \multicolumn{3}{|c|}{k-d forest}  \\
        \hline
        \# trees$^3$     & & &   1 &  8 & 64 & 512  \\
        \hline
        min depth        & & &    &  0 & 0 & 0  \\
        max depth / depth& & & 34 & 10 & 4 & 4 \\
        mean depth       & & &    & 5.2 & 0.9 & 0.1  \\
        var depth        & & &   & 2.1 & 1.1  & 0.5 \\
        \hline
        \hline
        evaluation [ms]       &-&-&  71 &  26 &  15 & 13 \\
        \hline
    \end{tabular}
    }
    \caption{
        Velocity field in a fluidized bed with 250 seeds.
    }
    \label{fig:lrvec2}
\end{figure}

\section{Vector field rendering}\label{sec:vecrender}
Vector fields are typically used to represent physical properties such as a magnetic field or flow velocity.
In the case of steady flow, i.e., a vector field that does not change with time, stream-, streak-, and pathlines coincide.
For a given vector field $\mathbf{f}:\R^3 \rightarrow \R^3$ a streamline with seed point $\mathbf{x}_0\in\R^3$ is given by the solution $\mathbf{x}(t), t\in\R^+$ of the initial value problem
\begin{equation}
    \label{eq:streamline}
    \begin{split}
        \dot{\mathbf{x}} = \mathbf{f}(\mathbf{x}),\\
        \mathbf{x}(0) = \mathbf{x}_0.
    \end{split}
\end{equation}
In the following we will assume that $\mathbf{f}$ is Lipschitz continuous, hence we have the existence of a unique local solution according to the theorem of Picard-Lindel\"of.
This assumption is fulfilled for LR-splines.
For the rendering of streamlines we employ a \emph{two-pass algorithm}, with a separate stage for computation and rendering.
In the rendering stage the local position within an element can be used for a lighting model, e.g., Blinn-Phong.
In addition, the tube can be colored for example by computing the local speed.
This means evaluating the LR-spline at the local position and computing the Euclidean norm.

\begin{figure}
    \centering
    \subfigure[
        Visualization of streamlines around a telescope, with color indicating speed from low (blue) to high (red).
        The rendering stage of the streamlines on a screen with resolution of $1280\times720$ takes about 4ms.
        ]
    {
        \includegraphics[width=.5\linewidth]{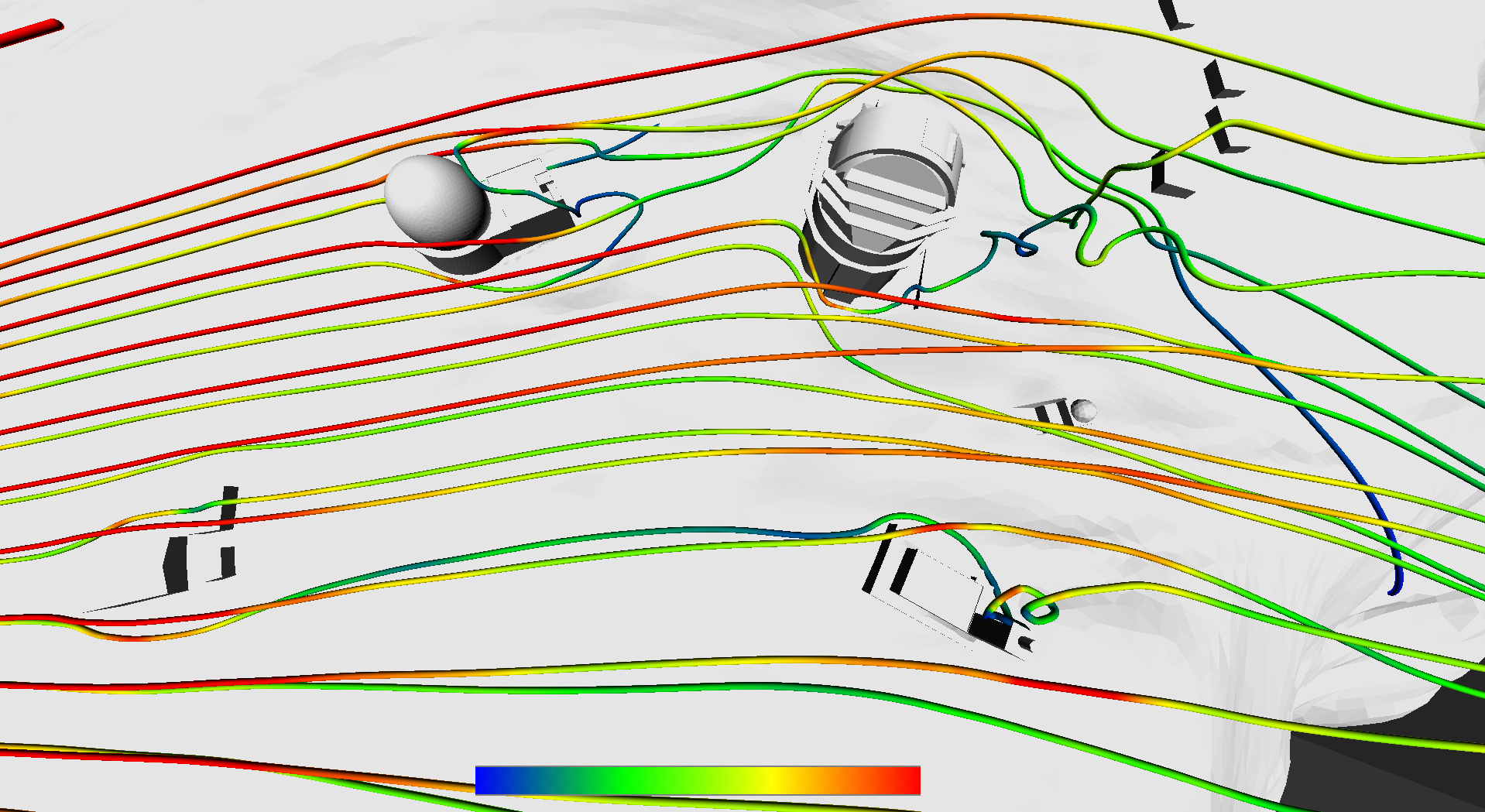}
        \includegraphics[width=.5\linewidth]{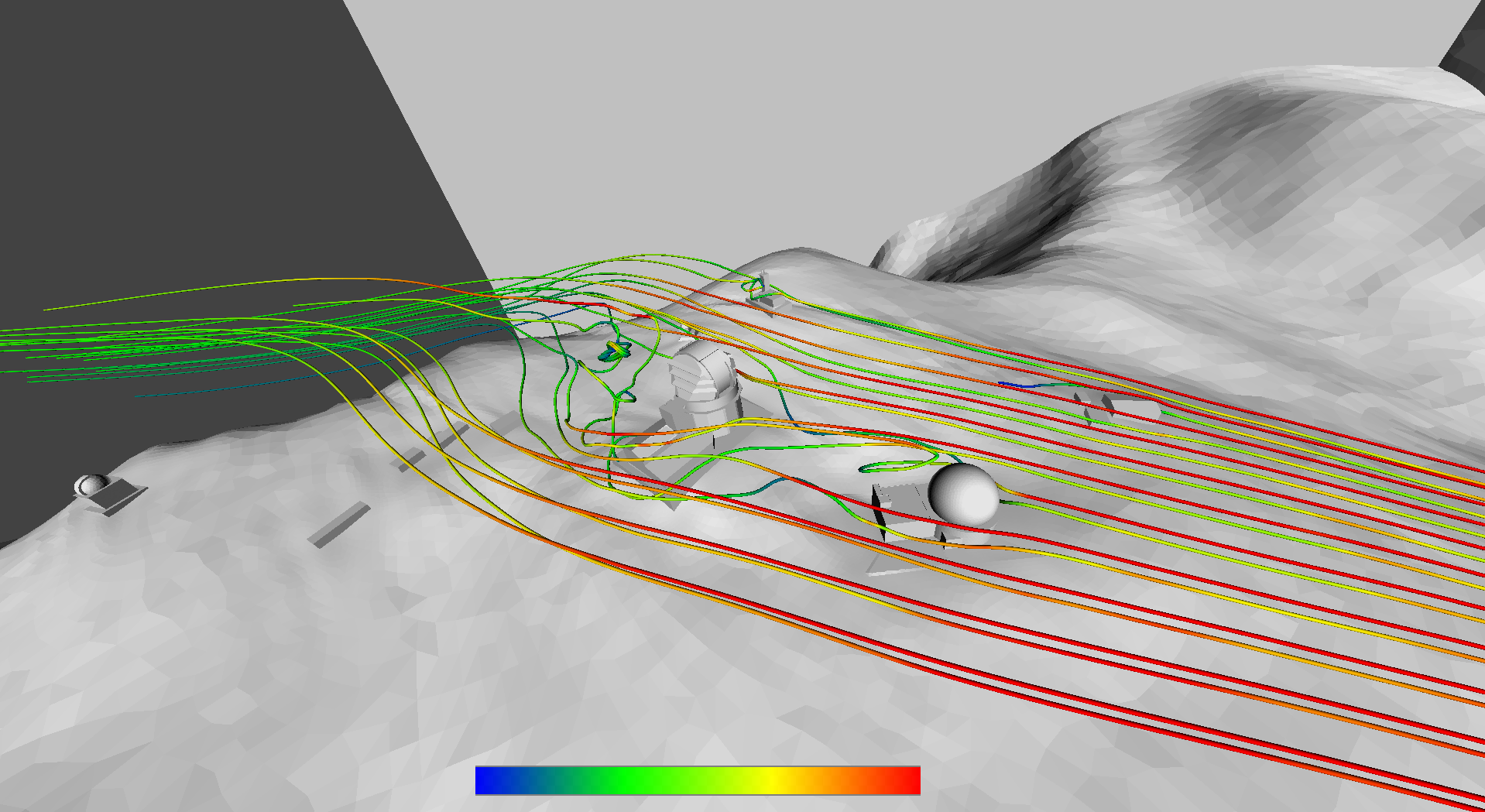}
    }
    \subfigure[
        Computation times for LR-spline evaluation for a representative approximation comparing different search methods.
        The measurement was performed on an NVIDIA Titan GPU.
        ]
    {
    \begin{tabular}{| l || c || c || c | c | c |}
        \hline
        \multicolumn{1}{|c||}{}& texture & octree & \multicolumn{3}{|c|}{k-d forest}  \\
        \hline
        \# trees$^3$     &   &    & 64  & 256 & 512    \\
        \hline
        min depth        &   &    & 0  &  0 &  0       \\
        max depth / depth&   & 15 & 49 & 29 &  20      \\
        mean depth       &   &    & 14 & 0.001 & 0.0002\\
        var depth        &   &    & 16 & 0.07 & 0.02   \\
        \hline
        \hline
        evaluation [ms] & - & 17 & 20 &  10 &   9     \\
        \hline
    \end{tabular}
    }
    \caption{
        Velocity field around a telescope with 88 seeds.
    }
    \protect\label{fig:lrvec1}
\end{figure}

\subsection{Background: computation of stream lines}\label{sec:compstream}
The computation of streamlines is done by discretizing the initial value problem \eqref{eq:streamline} with an explicit Runge Kutta (RK) method given by
\begin{equation}
    \label{eq:RK}
    \mathbf{x}_{n+1} = \mathbf{x}_n + h \sum_{i=1}^s b_i \mathbf{k}_i,
\end{equation}
where $h$ is the step size, $b_i$ are weights, and $\mathbf{k}_i=\mathbf{f}(\cdot)$ are increments at locations certain locations specific to the RK method.
We would like to point the reader to \cite{deuflhard2002scientific} for a good introduction to numerical methods for ODEs.
Note that in our case the evaluation of $\mathbf{f}$ means to evaluate an LR-spline.
The RK method is terminated if the streamline exits the LR-spline domain, a specified "time" is reached, or if the maximum number of samples is reached.

Typically, one seeks control over the error of the numerical approximation in order to ensure stability properties such as A-stability, see \cite{deuflhard2002scientific}.
This is achieved by basing the local step size on a method that produces a local error estimate, leading to an \emph{adaptive} method.
A popular choice are embedded RK methods that have two methods in the tableau, one of order $p$ and one of order $p-1$.

\subsection{Heuristic for adaptive sampling distance}\label{sec:adaptivesdd}
In case of vector fields given by an LR-spline, we propose an alternative to adaptive ODE solvers.
We propose a heuristic where the local step size $h$ naturally depends on the relative size of the LR-elements and the degree of the underlying spline.
To this end, we use the same local step size as described in \refsec{adaptivesd}.

\subsection{Implementation}
For the rendering of streamlines we employ a two-stage algorithm.
        \emph{Computation} of streamlines: This stage is implemented as a compute shader and triggered only when seeds are set and/or changed, and/or a new field is loaded.
        Standard (higher order) discretization methods are applied and the result is stored in an RGBA32F image buffer of the size: number of seed points times maximum number of discrete sampling points.
        The last component is used to indicate if a sample point is used or not (active).
        At the end of this and the next stage a memory barrier is set.
        \emph{Rendering} of the computed streamlines: For each segment that is stored and rendered active in the image buffer, a tube is rendered.
        For this tube a specific color map and lighting model is applied.
        The streamlines are visualized by rendering piecewise straight tubes, that are joined together by the discrete sample points $\mathbf{x}_n$ computed by the ODE solver in the compute stage.
        To this end we draw (number of seed points) x (maximum number of discrete sampling points) primitives (GL\_POINTS) and run a program consisting of the following shader stages:
        \begin{enumerate}
            \item \textbf{Vertex shader}: The primitives are used to trigger rendering and the vertex shader sets a random position in order to be consistent with the OpenGL spec.
           \item \textbf{Geometry shader}: The primitive ID (gl\_PrimitiveIDIn) is used to determine the seed number $s$, and sampling point $n$.
               From this, the points $\textbf{x}_n$, $\textbf{x}_{n+1}$ are extracted and used to compute and emit 4 points comprising the tube segment.
           \item \textbf{Fragment shader}: Finally, the local position within an element is determined and a lighting model, e.g., Blinn-Phong, is applied.
               The intersections, start and end segments are treated such that they appear round.
               In addition, the tube can be colored for example by computing the local speed.
               This means evaluating the LR-spline at the local position and computing the Euclidean norm.
        \end{enumerate}

Note, that it is possible to avoid having a fixed maximum number of discrete sample points per streamline by creating a linked list approach. We have chosen to use an image buffer with a fixed maximum number of discrete samples in order to achieve maximum performance.

\subsection{Results}\label{sec:vecresults}
In this section we provide examples showcasing interactive streamline computation and rendering.
We would like to point out that the repeated computation of streamlines -- although interactive computation for every frame is possible -- is only necessary when the vector field or the seed points are initiated or changed.

In this article we compare the following RK methods:
(RK1) explicit Euler method (1st order),
(RK2) midpoint method (2nd order),
(RK3) Kutta's 3rd order method,
(RK4) classic 4th order method,
(RK4 3/8) 3/8 rule (4th order),
(RKF5) Runge-Kutta-Fehlberg method (5th order used),
and the following \emph{adaptive/embedded} RK methods:
(HE) Heun Euler (2nd order),
(BS) Bogacki-Shampine (3rd order),
(RKF45) Runge-Kutta-Fehlberg method (5th order)
.

\begin{figure}
    \centering
    \subfigure[
        The single precision algoithm fails to converge, while the mixed precision converges to within single precision tolerance.
        ]{
        \includegraphics[width=0.7\linewidth]{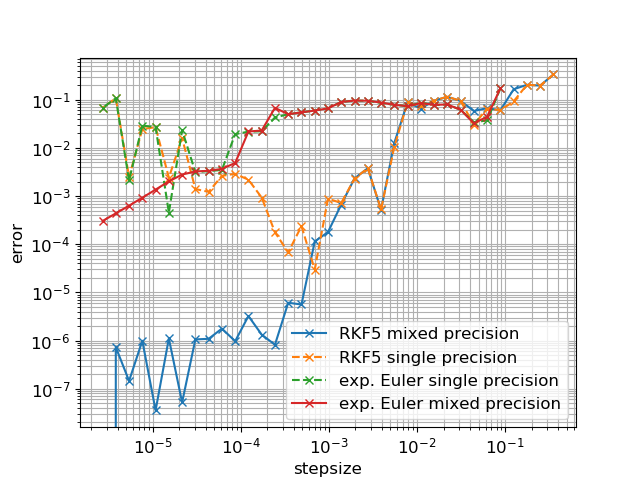}
    }
    \subfigure[
        There is a slight increase in computational time due of the mixed precision solver wrt. the single precision ones.
        ]{
        \includegraphics[width=0.7\linewidth]{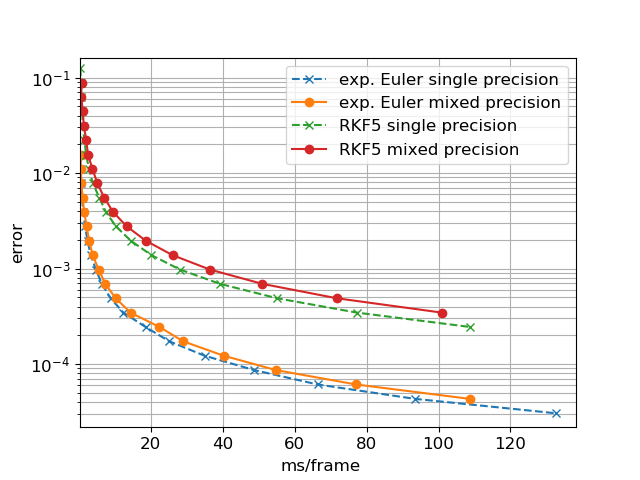}
    }
    \caption{
        Comparison of the results of the single and mixed precision algorithm with the error metric \refequ{vecerrormetric}.
        For clarity of description we illustrate the result for a few solvers, but the same basic behaviour is true for all solvers.
    }
    \label{fig:mixedprecision}
\end{figure}

\subsection{Element evaluation}
We start by presenting the case of particle velocity from a simulation of a fluidized bed in \reffig{lrvec2}.
The LR-spline of tri-degree (2,2,2) uses 5311 elements and 8 levels of refinement.
Neither the texture based, nor the octree based approach are applicable for element search, since there are knot values that cannot be written as multiples of $1/l^i$ where $l,i\in \N$.
Using the k-d forest based approach, the computation of 250 streamlines is performed with interactive frame rates, see \reffig{lrvec2}(b).
The fastest method is the kd-forest with $512^3$ trees.
The streamline rendering with tubes takes about 7ms for the view presented in \reffig{lrvec2}.

The pictures in \reffig{lrvec1} show screenshots of rendering of the simulation results of wind flow around a telescope.
The LR-spline with tri-degree (2,2,2) has in total 88872 elements and 2028 refinement levels are used.
Regarding the different methods for element look-up, the texture based approach is not applicable due to size of the smallest element, i.e., it would require a larger texture than OpenGL allows.
\reffig{lrvec1}(b) shows that the algorithm based on an octree is faster than a k-d forest based approach with $64^3$ trees.
However, as the number of trees is increased, the k-d forest based approach becomes faster.

\subsection{Mixed precision}
Next, we would like to point out that the case of wind flow around a telescope \reffig{lrvec1} requires very small step sizes (due to very small elements owing to 2082 levels of refinement).
This leads to the fact that the RK solvers in \emph{single} precision do not converge under refinement, see \reffig{mixedprecision}.
However, on consumer GPUs switching to double precision is up to 32 times slower.
We therefore implemented a \emph{mixed precision} solver, where the outer loop in \refequ{RK} is performed in double precision, and the spline evaluation remains in single precision.
\reffig{mixedprecision} shows a slight increase in computational time.

\subsection{Efficiency}
In the end, the most efficient method, wheather adaptive or heuristic, should be preferred.
In order to be objective, we define the following error metric for $J$ number of streamlines
\begin{equation}
    \label{eq:vecerrormetric}
    E (\{y_j(t^i), \scriptstyle 1\leq j\leq J, 1\leq i \leq I \textstyle \}) = \underset{j\in{1,..,J}}{\max} \left\{ \sum_{i=1}^I \Delta t^i \|x_j(t^i) - y_j(t^i)\|_{L^2} \right\},
\end{equation}
where $x_j$ is the analytical solution of \refequ{streamline} and $y_j(t^i)$ is a discrete approximation at the points $t_j$.
In the absence of an analytical solution, we approximate $x_j(t)$ by computing a reference solution with the fifth order RKF5 solver using a very small (fixed) step size.

\begin{figure}
    \centering
    \subfigure[
        Adapitve methods are slightly more efficient as the proposed heuristic for different tolerances (adaptive methods) and base step sizes (heuristic methods).
        However, it is important to note that the adaptive methods fail to reach errors of $1e-4$ or lower all stream lines.
        ]{
        \includegraphics[width=0.7\linewidth]{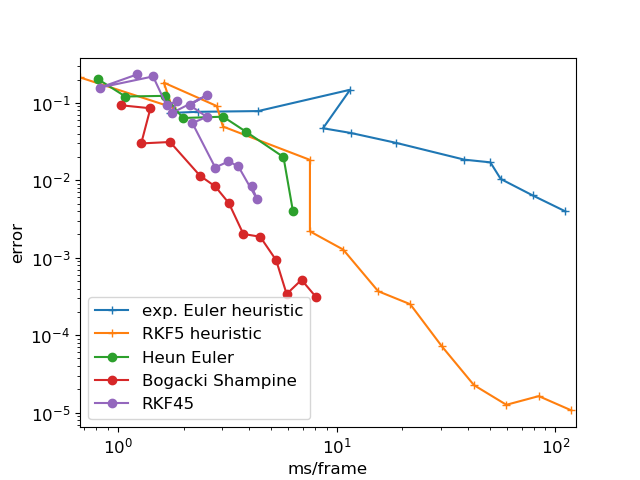}
    }
    \subfigure[
        Local step size of the methods.
        The general bahaviour of the heuristic and the adaptive methods is the same.
        ]{
        \includegraphics[width=0.7\linewidth]{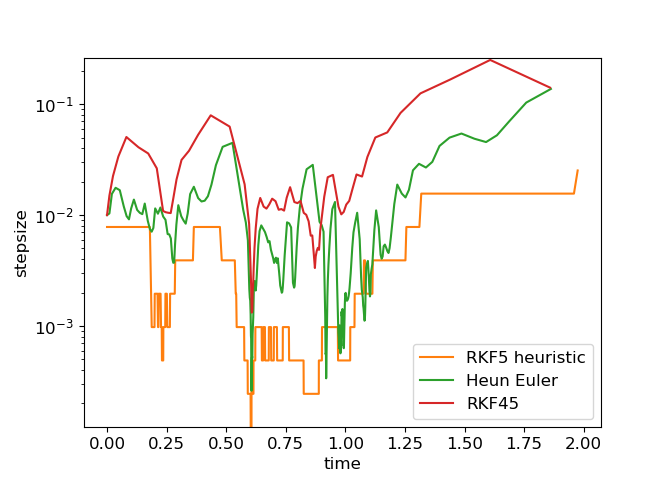}
    }
    \caption{
        Comparison of the adaptive and the heuristic solvers for the case shown in \reffig{lrvec1}.
        The error metric is defined in \refequ{vecerrormetric}.
    }
    \label{fig:vec_efficiency}
\end{figure}

\reffig{vec_efficiency}(b) shows that the adaptive and the suggested heuristic method use very similar local step sizes for streamlines in the case shown in \reffig{lrvec1}.
Judging from \reffig{vec_efficiency}(a), the standard adaptive (embedded) RK methods are slightly more efficient than the heuristic methods.
However, it seems that the heuristic method leads to a more stable outcome, as the method converges under refinement of the base step size.
Although most (approximations of) streamlines the adaptive methods converge as well, there are a few where it seems impossible to reach an error less than $\mathcal{O}(1e-4)$.
Note, that the embedded methods only control the \emph{local} error, but we are interested in the global error, as defined in \refequ{vecerrormetric}.
Note also, that the base method of RKF45 and RKF5 heuristic is exactly the same, as well as 

\section{Conclusion}\label{sec:conclusion}
A novel, interactive method for volume visualization and streamline rendering for LR-splines has been presented.
Overall, the best acceleration structure for quick element look-up depends on the specific case.
For examples with little refinement and restrictions on the knots values texture look-ups are fastest.
In case of many levels of refinement and general knot placement a k-d forest acceleration structure becomes indispensable.
A heuristic for adaptive step sizes based on the underlying LR-structure is presented.
Overall, interactive frame rates are demonstrated.
In the future we plan extend our algorithm to Isogeometric Analysis results and vector field rendering with LIC.
Another paper will focus on multilevel B-spline approximation (MBA) algorithm.

\section*{Acknowledgements}
The research leading to these results has received funding from the European Community's Seventh Framework Programme (FP7-ICT-2013-11) in the scope of the VELaSSCo project under Grant Agreement Number 619439.
MR brain data courtesy Siemens Medical Systems, Inc., Iselin, NJ and University of North Carolina.
DEM data (Fluidized Bed) courtesy of the School of Engineering at the University of Edinburgh.
FEM data (Telescope) courtesy the GiD and Kratos groups at CIMNE (the International Center for Numerical Methods in Engineering).

\bibliographystyle{cag-num-names}
\bibliography{LRVolume}

\end{document}